\documentstyle[12pt]{article}
\newlength{\pubnumber} 

\catcode`\@=11
\@addtoreset{equation}{section}
\def\section{\@startsection{section}{1}{\z@}{3.5ex plus 1ex minus .2ex}
 {2.3ex plus .2ex}{\large\bf}}
\def\subsection{\@startsection{subsection}{2}{\z@}{2.3ex plus .2ex}
 {2.3ex plus .2ex}{\bf}}

\begin{document}

\begin{titlepage}
\samepage{
\setcounter{page}{1}
\rightline{UFIFT-HEP-96-9}
\rightline{\tt hep-ph/9603272}
\rightline{March 1996}
\vfill
\begin{center}
 {\Large \bf New Dark Matter Candidates \\
     Motivated From Superstring Derived Unification\\}
\vfill
 {\large Sanghyeon Chang\footnote{
   E-mail address: schang@phys.ufl.edu},
   Claudio Corian\`{o}\footnote{
   E-mail address: coriano@phys.ufl.edu}
   $\,$and$\,$ Alon E. Faraggi\footnote{
   E-mail address: faraggi@phys.ufl.edu}\\}
\vspace{.12in}
 {\it  Institute For Fundamental Theory, \\
  Department of Physics, \\
  University of Florida, \\
  Gainesville, FL 32611 USA\\}
\end{center}
\vfill
\begin{abstract}
  {\rm
Perturbative gauge coupling unification in realistic superstring models
suggests the existence of additional heavy down--type quarks, beyond the
minimal supersymmetric standard model. The mass scale
of the heavy down--type quarks is constrained by requiring agreement
between the measured low energy gauge parameters and the string--scale
gauge coupling unification. These additional quarks arise and may be
stable due to the gauge symmetry breaking by ``Wilson lines'' in the
superstring models. We argue that there is a window in the parameter
space within which this down--type quark is a good candidate for the
dark matter.
}
\end{abstract}
\vfill
\smallskip}
\end{titlepage}

\setcounter{footnote}{0}

\def\beq{\begin{equation}}
\def\eeq{\end{equation}}
\def\beqn{\begin{eqnarray}}
\def\eeqn{\end{eqnarray}}
\def\AEF{A.E. Faraggi}
\def\NPB#1#2#3{{\it Nucl.\ Phys.}\/ {\bf B#1} (19#2) #3}
\def\PLB#1#2#3{{\it Phys.\ Lett.}\/ {\bf B#1} (19#2) #3}
\def\PRD#1#2#3{{\it Phys.\ Rev.}\/ {\bf D#1} (19#2) #3}
\def\PRL#1#2#3{{\it Phys.\ Rev.\ Lett.}\/ {\bf #1} (19#2) #3}
\def\PRT#1#2#3{{\it Phys.\ Rep.}\/ {\bf#1} (19#2) #3}
\def\MODA#1#2#3{{\it Mod.\ Phys.\ Lett.}\/ {\bf A#1} (19#2) #3}
\def\IJMP#1#2#3{{\it Int.\ J.\ Mod.\ Phys.}\/ {\bf A#1} (19#2) #3}
\def\nuvc#1#2#3{{\it Nuovo Cimento}\/ {\bf #1A} (#2) #3}
\def\etal{{\it et al,\/}\ }
\hyphenation{su-per-sym-met-ric non-su-per-sym-met-ric}
\hyphenation{space-time-super-sym-met-ric}
\hyphenation{mod-u-lar mod-u-lar--in-var-i-ant}


\setcounter{footnote}{0}

Substantial observational evidence indicates
that most of the mass in the universe is invisible.
The determination of the nature of this
``dark matter'' is one of the most important challenges confronting
modern physics. In this paper we study the
possibility that the dark matter is composed of heavy QCD
color triplets that arise in standard--like superstring models.
These are regular down--like heavy quarks with the
down--type charge assignment. The existence of such heavy quarks
is motivated from string--scale gauge coupling unification \cite{DF}. Due
to its role in the string unification, we refer to this type
of particle as the uniton.
The uniton can be stable due to the breaking of gauge symmetries
by ``Wilson lines'' in superstring models.
The additional down--like quarks are obtained in the superstring models
from sectors that arise due to the ``Wilson line'' breaking.
As a result they acquire ``fractional'' charges under the $U(1)_{Z^\prime}$
gauge symmetry while the Standard Model states have the standard
$SO(10)$ charge assignment. The stability of the uniton
results from a gauge $U(1)$ symmetry, which is left unbroken
down to low energies, or from a local discrete symmetry \cite{KW}.
It forms bound heavy meson states
with the Standard Model down and up quarks.
We estimate the electromagnetic
mass difference between the charged and neutral heavy
U--meson states and argue that, over some region of the parameter space,
the neutral meson state is the lighter one. We estimate
the contribution of the uniton to the relic density.
We examine other astrophysical and terrestrial bounds on a stable
uniton, and propose that there exist a window in the parameter space
in which the uniton is a good dark matter candidate.
In this paper we shall present our main results and
further details of the analysis will be given in ref. \cite{bigpaper}.

In their low energy limit, heterotic string theories give rise to
$N=1$ supersymmetry. While other possible extensions of the
Standard Model are highly constrained or ruled out by experiments,
supersymmetric theories are in agreement with the available data.
The attractive motivation for supersymmetric theories
is not without flaw. While the unification of the
gauge coupling in the Minimal Supersymmetric Standard Model (MSSM)
occurs at a scale $M_{\rm MSSM}\approx2\times10^{16}~{\rm  GeV}$,
string theories predict a larger unification scale \cite{ginsparg},
typically $M_{\rm string}\approx g_{\rm string}\times5\times10^{17}$~GeV
where $g_{\rm string}\approx0.8$ at the unification scale.
Thus, an order of magnitude separates the MSSM and string unification scales.

It would seem
that in an extrapolation of the gauge parameters over
fifteen orders of magnitude, a problem involving a single
order of magnitude would have many possible resolutions.
Indeed, in superstring models there are a priori many
possible effects that can account for the discrepancy.
Surprisingly, however, the discrepancy is not easily resolved.
The validity of string gauge coupling unification
must be examined in the context of realistic string models.
The superstring models in the free fermionic formulation
represent a class of phenomenologically appealing models.
Not only do these models naturally yield three generation
with a plausible fermion mass spectrum, but perhaps of equal
importance is the fact that these models predict $\sin^2\theta_W=3/8$
at the string unification scale. This rather common result from the
point of view of regular GUT models is highly non trivial from the
point of view of string models.
In ref. \cite{DF}
it was shown, in a wide range of realistic
free fermionic models, that heavy string threshold corrections,
non-standard hypercharge normalizations,
light SUSY thresholds or intermediate
gauge structure, do not resolve the problem.
Instead,
the problem may only be resolved due to the existence of
additional intermediate matter thresholds, beyond the MSSM
\cite{Gaillard,GCU}. This additional matter takes the form of
color triplets and electroweak doublets,
in vector--like representations. Remarkably, some string models
contain in their massless spectrum the additional states
with the specific weak hypercharge assignments,
needed to achieve string scale unification \cite{GCU}.
Possible scenarios to generate the needed mass scales
from the string models have been discussed in the literature.

A model in the free fermionic formulation is
generated by a consistent set of boundary condition basis
vectors \cite{FFF}. The physical spectrum
is obtained by applying the generalised GSO projections.
The first five basis vectors in our models consist of the
so--called ``NAHE--set''
$\{{\bf 1}, S,b_1,b_2,b_3\}$ \cite{nahe}.
At the level of the NAHE set
the gauge group is $SO(10)\times SO(6)^3\times E_8$, with 48 generations.
The number of generations is reduced to three and the $SO(10)$
gauge group is broken to $SU(3)\times SU(2)\times U(1)^2$
by adding to the NAHE set three additional basis vectors,
$\{\alpha,\beta,\gamma\}$. The basis vectors $\alpha$ and
$\beta$ break the $SO(10)$ symmetry to $SO(6)\times SO(4)$.
The basis vector $\gamma$ breaks the $SO(2n)$ symmetries to
$SU(n)\times U(1)$.
It is useful to note the correspondence between free fermionic models
and orbifold models. The free fermionic models correspond to
toroidal $Z_2\times Z_2$ orbifold models with nontrivial
background fields. The Neveu--Schwarz sector corresponds to the
untwisted sector, and the sectors $b_1$, $b_2$ and $b_3$ correspond to the
three twisted sectors of the orbifold models.
The three sectors which break the $SO(10)$ symmetry
correspond to Wilson lines in the orbifold terminology.

The massless spectrum of the superstring standard--like models
consists of three 16 representations of $SO(10)$ from the sectors
$b_1$, $b_2$ and $b_3$ decomposed under the final
gauge group.
The Neveu--Schwarz sector produces three pairs of electroweak doublets
and several $SO(10)$ singlet fields. The sector $b_1+b_2+\alpha+\beta$
produces one or two additional electroweak doublet pairs and $SO(10)$
singlet fields.
Additional massless states are obtained from sectors
that arise from combinations of the vectors $\{\alpha,\beta,\gamma\}$
with the vectors of the NAHE set. The sectors $b_j+2\gamma$ $(j=1,2,3)$
produce three 16 representations of the hidden $SO(16)$ gauge subgroup
decomposed under the final hidden gauge group.
All the states above can either fit into $SO(10)$ multiplets
or are $SO(10)$ singlets.

The massless spectrum of the superstring models
contain additional massless states that
do not fit into $SO(10)$ multiplets. For example, in the model
of ref. \cite{GCU} the sector $1+\alpha+2\gamma$ produces a pair
of color triplets with quantum numbers $(3,1,1/6)$ and $({\bar 3},1,-1/6)$
under $SU(3)\times SU(2)\times U(1)_Y$. The sectors
$1+b_{1,2,3}+\alpha+2\gamma$ produce three pairs of electroweak doublets
with charges $(1,2,0)$. Such states are obtained
from sectors that break the $SO(10)$ symmetry to $SO(6)\times SO(4)$
and therefore also appear in the $SO(6)\times SO(4)$
superstring models \cite{alr}.
These states carry fractional electric charge
and cannot be candidates for dark matter, as there are strong
constraints on their possible mass scale and abundance.

In addition to the states above there is an additional class
of massless states that are more interesting from cosmological
considerations. These states are regular down--like quarks with the
down--type charge assignments. These states are obtained
from sectors that break
the $SO(10)$ symmetry to $SU(3)\times SU(2)\times U(1)^2$.
For example, in the model
of ref. \cite{GCU} the sectors $b_1+b_{2,3}+\beta\pm\gamma$ produce
two pairs of color triplets with charges $(3,1,1/3)$ and
$({\bar 3},1,-1/3)$. These states arise in the string models
from sectors that break the $SO(10)$ symmetry. They carry fractional
charges under the $U(1)_{Z^\prime}$ symmetry, which
is embedded in $SO(10)$ and is orthogonal to $U(1)_Y$.
Consequently, these color triplets cannot fit into $SO(10)$ multiplets.
This property enables the stability of this type of color
triplets. To examine whether these states can decay into the
Standard Model states we must examine
their interactions. For example, examination of the superpotential
terms of the model of ref. \cite{eu} shows that such interactions
terms are obtained if the $U(1)_{Z^\prime}$ symmetry is broken.
The near stability of the uniton in this case can be associated with
the existence of a low energy $Z^\prime$ gauge boson. However,
this need not be the case.
Analysis of the nonrenormalizable terms in the model of ref. \cite{GCU}
shows that superpotential terms between the uniton and the
standard model states are not generated at any order \cite{ps}.
In this model the uniton cannot decay into the standard model states
even if the $U(1)_{Z^\prime}$ symmetry is broken.
Therefore, in this model the uniton is stable.
In this model the stability of the uniton may be associated
with a local discrete symmetry \cite{KW}.

The uniton forms bound meson states with the up and down quarks.
In order to estimate the mass splitting of these two heavy mesons,
 we borrow
from recent work of ref. \cite{LS} which gives a
general formula
for the mass splitting of two heavy-light mesons ($Q\bar{u}, Q\bar{d}$).
We set $\Delta M = M^- - M^0$ to be the mass difference between the
 negatively charged meson and the zero charged one respectively.
Intuitively, there are 2 contributions of opposite sign which
contribute to the
splitting :
the difference in mass between the two constituent quarks,
and the electromagnetic splitting.
The basic strategy \cite{LS}
is to relate the order $e^2$ isospin breaking
corrections to the forward Compton
scattering amplitude, $T$. The heavy quark effective theory,
in the large $N_c$ limit, is then used to write $T$ in terms
of heavy meson form factors by introducing a dispersion relation.
The largest contribution to the
spectral function comes from the
lowest-lying single heavy meson states,
while the continuum contribution is negligible.
The infinite sum over the intermediate states
is truncated after the first few terms, and includes,
beside the contribution of the first excitation of the heavy meson,
also a minimal sets
of light mesons $\rho,$ $\rho'$ $\omega$ and $\omega'$.
These are introduced
in order to obtain consistency (at large $N_c$) with the asymptotic
behaviour of the form factors from the heavy quark theory,
as expected from the dimensional counting rules.

In the case of our heavy meson system the ${\cal O}(1/m_Q^2)$ corrections
are negligible. The electromagnetic contribution to the mass difference, up
to 30\% accuracy, is \cite{LS}
\begin{eqnarray}
\left. M_{U^+}-M_{U^0}\right|_{EM}\sim&& +1.7 \nonumber\\
-0.13&&\left({\beta\over 1 {\rm GeV}^{-1}}\right)
-0.03\left({\beta\over 1 {\rm GeV}^{-1}}\right)^2,
\label{one}
\end{eqnarray}
where $\beta\sim 1/m_Q$ measures the matrix element of the decay of the
first excited heavy meson state into the ground state plus a photon.
Using the values from the particle data book for $m_u$, $m_d$ and
$m_u/m_d$ \cite{expalpha} we obtain $M_{U^+}-M_{U^0}\sim1$ MeV.
The result given in eq.~ (\ref{one}) does not include the contribution to
the mass difference coming from $SU(2)$ splitting of the two constituent
quarks inside the two mesons nor the strong interactions effects.
An exact calculation of this splitting is not possible with our present
understanding of low energy QCD. It is not unplausible that
the electromagnetic mass difference and the
strong interaction effects are of the
order of a few MeV and can overcompensate the mass difference due to the two
constituent quarks in the two mesons, therefore making
$U^+$ heavier than $U^0$.
It is plausible that the splitting between $U^+$ and $U^0$ behaves similarly
to the B-meson case, in which the mass difference between $B^+$ and $B^0$
is comparable to zero at the one sigma confidence level. We refer to
\cite{bigpaper} for a more detailed discussion of this issue.
We therefore conclude that the possibility of having
a neutral heavy meson lighter than the charged one is not ruled out.

We now discuss the cosmological and astrophysical bounds.
The uniton is a strongly interacting particle and therefore it remains
in  thermal equilibrium until it becomes nonrelativistic.
To calculate the relic density of the uniton we need to know its
decoupling temperature from the thermal bath.
In the nonrelativistic limit, $T/M<1$, the uniton annihilation rate
is given by
\begin{eqnarray}
\Gamma =
 <\sigma|v|> n_{eq}
\simeq\frac{\pi N \alpha_s^2 }{M^2 } n_{eq},
\end{eqnarray}
where $M$ is the mass of the uniton, $\alpha_s$ is the strong coupling at
decoupling, $n_{eq}$ is the number density of the uniton at
equilibrium. $N$ is a summation over all the available annihilation
channels and is given by
$N=\sum_f{a_f}$.
The amplitudes $a_f$ are obtained by calculating the annihilation cross
section of the uniton to all the strongly interacting particles,
which include the six flavors of quarks and squarks and the gluons and
the gluinos. The final states are taken to be massless.
We obtain $a={4/3}$ for quarks; $a=14/27$ for gluons;
$a=2/3$ for squarks and $a=64/27$ for gluinos.

The uniton decouples from the thermal bath when its annihilation rate
falls behind the expansion rate of the universe.
In the expanding universe,
the evolution equation of the particle density in comoving volume
is
\begin{equation}
\frac{dY}{dx} = -\lambda x^{-2} (Y^2-Y^2_{eq}).
\end{equation}
Here $Y=n/s$, $x\equiv M/T$ and
\begin{eqnarray}
\lambda&=&\left.\frac{x<\sigma|v|> s}{H}\right|_{x=1}
=0.83 N\alpha_s^2\frac{g_*s}{\sqrt{g_*}}\frac{m_{pl}}{ M} .
\end{eqnarray}
Here the entropy $s$ is $(2\pi^2/45) g_{*s} m^3 x^{-3}$.
The decoupling condition $dY/dx \simeq 0$ gives \cite{KTB}
\begin{equation}
x_{dec} = \ln [(2+c)\lambda ac] -\frac{1}{2}\ln\{\ln [(2+c)\lambda ac]\},
\end{equation}
where $ a= 0.145(g/g_{*s})$ and $c$ is $Y(T_{dec})/Y_{eq}(T_{dec})$, which is
of
order one.
We approximately estimate the decoupling temperature to be of the form
\begin{equation}
T_{dec}\simeq\frac{M}{\ln(m_{pl}/M)}.
\end{equation}

The uniton density at the present universe is
\begin{equation}
Y_0=\frac{3.79 x_{dec}}{\sqrt{g_*} m_{pl} M <\sigma|v|>},
\end{equation}
where we set $g_* =g_{*s}$, since the decoupling temperature is high.
Since the relic energy density of a massive decoupled particle is
$\rho_0=M s_0 Y_0$,
we can estimate the ratio of energy density to the critical energy density
at the present universe to be
\begin{eqnarray}
\Omega_0 h^2 \equiv \frac{\rho h^2}{\rho_c}
\simeq 10^9 \frac{\ln(m_{pl}/M) M^2}
{N\alpha_s^2\sqrt{g_*}m_{pl}}{\rm GeV}^{-1}.
\end{eqnarray}
The cosmological data indicates that $0.1 <\Omega h^2<1 $.
Using this condition we get an upper bound on the mass of the uniton
\begin{equation}
M <10^5 \alpha_s \left(N\sqrt{g_*}\ln(m_{pl}/M)\right)^{1/2}\mbox{GeV}.
\end{equation}
We now discuss the case with
inflation and the decoupling
temperature is greater than the reheating temperature.
If the reheating temperature $T_R$ is smaller than the decoupling temperature
$T_{dec}$, those particles will be diluted away and regenerated
after reheating by out-of-equilibrium production.
Since the uniton is completely diluted after the inflation,
the relic density at the reheating temperature is 0.
We can approximate it as
\begin{equation}
\frac{dY}{dx} =  \lambda x^{-2}Y_{eq}^2 \label{boltzman2},
\end{equation}
with $Y_{eq}= 0.145 g/g_* x^{3/2} e^{-x}$.
Integrating this relation from the reheating temperature to the present
temperature
we get
\begin{equation}
Y_0= \frac{\lambda g^2}{2}
 \left(\frac{0.145}{g_*}\right)^2 \left(x_r+\frac{1}{2}\right) e^{-2x_r},
\end{equation}
where $x_r\equiv M/T_R$. $T_R$ is the reheating temperature, and
\begin{equation}
\Omega_0h^2 \simeq 9\times 10^3 N\alpha_s^2 g^2\frac{m_{pl}}{M}
\left(\frac{200}{g_*}\right)^{1.5} \left(x_r+\frac{1}{2}\right) e^{-2x_r}.
\end{equation}
We can estimate the bound on the mass,
\begin{equation}
M>T_R\left[25+\ln(\sqrt{M/T_R})\right].
\end{equation}
Without inflation, we have  a strict bound on the mass of the uniton,
which is around $10^5$ GeV.
Inflation can raise the mass bound to any arbitrary order,
depending upon the estimated value of the reheating temperature.

We remark that there are 3 windows $W1$,
$W2$ and $W3$ for strongly interacting dark
matter \cite{SGED90} which possibly meet our requirements.
In W1 we need 10 GeV $< M < 10^4$ GeV
and a scattering cross section of the heavy hadron to
proton about $10^{-24}\sim 10^{-20}$cm$^2$.
In W2 and W3 we need $10^5$ GeV $< M <10^7$ GeV
and $M>10^{10}$ GeV, respectively,
 assuming a cross section less than
$10^{-25} $cm$^2$.
A charged bound state will form a hydrogen-like atom and  have a
cross section about $\sim 10^{-16}$cm$^{-2}$.
The upper bound on the cross section is fixed by the neutron star
lifetime \cite{GDR90}.
Note that there are no limits from nucleosynthesis on a
nonrelativistic heavy uniton because of its low density relative to
proton and neutron and its low energy density compared with the energy density
of radiation at nucleosynthesis.
COBE data fits well with $\Omega_0 h\simeq 0.25$ cold dark matter model
or mixed dark matter model with 20\% of cold dark matter.
The heavy stable uniton is a good candidate of the cold dark matter
component. We conclude that the uniton can evade all the currently available
experimental constraints. We therefore propose that the uniton is
a good dark matter candidate.

The consistency of perturbative
string unification with low energy data seems to require the
existence of additional intermediate matter thresholds \cite{witten}.
Remarkably, the same states
that are required for the string scale unification,
solve at the same time the dark matter problem. As there exist
many possible scenarios for the scales of the additional matter
states \cite{DF}, the string scale unification constraints can
be compatible with the constraints on the uniton dark matter \cite{bigpaper}.
In the free fermionic standard--like models the additional matter
states are obtained from sectors that correspond to ``Wilson lines''
in orbifold models. It is well known that the ``Wilson line'' breaking
in superstring models results in physical states with fractional electric
charge \cite{ww}. Due to electric charge conservation, fractionally charged
states are stable. As there exist
strong constraints on their masses and abundance,
fractionally charged states cannot constitute the dark matter.
Such states must be diluted away or extremely massive.
Remarkably, however, the same ``Wilson line'' breaking mechanism,
that produces matter with fractional electric charge, is also responsible for
the existence of
states which carry the ``standard'' charges under the Standard Model
gauge group but carry fractional charges under the $U(1)_{Z^\prime}$
symmetry. In the free fermionic standard--like models the three light
generations are obtained from the three 16 representations of $SO(10)$.
Consequently, due to the $U(1)_{Z^\prime}$ charge conservation,
the additional matter states cannot decay into the standard model states.
This may be the case even after $U(1)_{Z^\prime}$ symmetry breaking, in which
case a local discrete symmetry is left unbroken.
It is very encouraging, in our opinion, that the stability of the
uniton is associated with a gauge symmetry or a local discrete symmetry.
As global symmetries are in general expected to be violated by
quantum gravity effects, this fact is an important advantage over some
other dark matter candidates.
Due to the general applicability of the ``Wilson line'' breaking mechanism
in superstring models, the uniton may in fact be generic to string models
that aim at obtaining the standard model gauge group directly at
the string scale.
It will be of further interest to study additional
cosmological and phenomenological implications that this type of matter
might have. Such work is in progress.


We thank G. Bodwin, R. Field,
D. Kennedy, P. Ramond, P. Sikivie and C. Thorn
for discussions.
This work was supported in part by DOE Grant No.\ DE-FG-0586ER40272
and by KOSEF.

\bibliographystyle{unsrt}

\end{document}